\documentclass[conference]{IEEEtran}
\IEEEoverridecommandlockouts
\usepackage{cite}
\usepackage{amsmath,amssymb,amsfonts}
\usepackage{graphicx}
\usepackage{textcomp}
\usepackage{xcolor}
\usepackage{booktabs}
\usepackage{subfigure}
\usepackage{algorithm}
\usepackage{algpseudocode}
\usepackage{amsmath,amssymb} % para símbolos matemáticos
\usepackage{microtype}      % opcional, melhora a tipografia
\usepackage{tikz}
\usepackage{graphicx}
\usepackage[nolist]{acronym}
\usepackage{fontawesome}
\usepackage{tabularx}
\usepackage{url}
\usepackage{xurl}
\usetikzlibrary{fit,positioning,calc,arrows.meta}
\usetikzlibrary{arrows.meta, positioning, calc, shapes.geometric, fit, backgrounds}

\def\BibTeX{{\rm B\kern-.05em{\sc i\kern-.025em b}\kern-.08em
    T\kern-.1667em\lower.7ex\hbox{E}\kern-.125emX}}
\begin{document}

%\title{An Agentic-based Control-Loop for Stable QoS Enforcement in Mobile Networks}
%\title{Latent Space Based Data Plane Reservation for Critical Network Slicing in Mobile Networks}
%\title{Data Plane Reservation Using Latent Space for Critical Network Slicing in Mobile Networks}
%\title{Data Plane Reservation Using Latent Space for Critical Network Slicing}
%\title{Latent-State Reservation for Critical Consumer Traffic in Network Slicing}
\title{Conflict-Aware Autonomous Resource Orchestration for 5G Edge Slicing Using Multi-Agent Systems and LLMs}

\author{\IEEEauthorblockN{Nathan R. Prediger\textsuperscript{1}, Rodrigo Moreira\textsuperscript{2}, Juliano Araujo Wickboldt\textsuperscript{1},\\ Tereza C. Carvalho\textsuperscript{3},
Pedro Oliveira\textsuperscript{4},
Paulo Novais\textsuperscript{4},
Flávio de Oliveira Silva\textsuperscript{4}}
\IEEEauthorblockA{
\textsuperscript{1}Federal University of Rio Grande do Sul (UFRGS), Porto Alegre, Brazil\\
\textsuperscript{2}Federal University of Viçosa (UFV), Minas Gerais, Brazil\\
\textsuperscript{3}University of São Paulo (USP), São Paulo, Brazil\\
\textsuperscript{4}University of Minho (UMinho), Braga, Portugal\\
Emails: nrprediger@inf.ufrgs.br, rodrigo@ufv.br, jwickboldt@inf.ufrgs.br, terezacarvalho@usp.br,\\ pedro.jose.oliveira@algoritmi.uminho.pt,\{pjon, flavio\}@di.uminho.pt}
}

\maketitle
%Numbering
\acrodef{3GPP}{3rd Generation Partnership Project}
\acrodef{5G}{5th Generation Mobile Network}
\acrodef{6G}{6th Generation Mobile Network}
%-----A-----
\acrodef{AI}{Artificial Intelligence}
\acrodef{API}{Application Programming Interface}
\acrodef{AI4Net}{\ac{AI} for Networking}
\acrodef{AIDER}{Aerial Image Dataset for Emergency Response}
\acrodef{AMF}{Access and Mobility Management Function}
\acrodef{AIaaS}{Artificial Intelligence-as-a-Service}
\acrodef{AGORA}{Agentic Green Orchestration Architecture for Beyond 5G Networks}
\acrodef{IID}{Independent and Identically Distributed}
%-----B-----
\acrodef{B5G}{Beyond Fifth Generation}
\acrodef{BESS}{Berkeley Extensible Software Switch}
\acrodef{BPF}{Berkeley Packet Filter}
%-----C-----
\acrodef{CBR}{Constant Bit Rate}
\acrodef{CSV}{Comma-Separated Values}
\acrodef{CPU}{Central Processing Unit}
\acrodef{CNN}{Convolutional Neural Network}
\acrodef{CNNs}{Convolutional Neural Networks}
%-----D-----
\acrodef{DoS}{Denial of Service}
\acrodef{DDoS}{Distributed Denial of Service}
\acrodef{DDPG}{Deep Deterministic Policy Gradient}
\acrodef{DNN}{Deep Neural Network}
\acrodef{DPDK}{Data Plane Development Kit}
\acrodef{DRL}{Deep Reinforcement Learning}
\acrodef{DT}{Decision Tree}
\acrodef{DQN}{Deep Q-Network}
%-----E-----
\acrodef{ETSI}{European Telecommunications Standards Institute}
\acrodef{eNWDAF}{Evolved Network Data Analytics Function}
\acrodef{EIF}{Energy Information Function}
\acrodef{E2E}{End-to-End}
%-----F-----
\acrodef{FIBRE}{Future Internet Brazilian Environment for Experimentation}
\acrodef{FPR}{False Positive Rate}
\acrodef{FAR}{Forwarding Action Rule}
%-----G-----
\acrodef{GNN}{Graph Neural Networks}
\acrodef{GPU}{Graphics Processing Unit}
\acrodef{GTP}{GPRS Tunnelling Protocol}
\acrodef{GTP-U}{GPRS Tunneling Protocol User Plane}
%-----H-----
\acrodef{HTM}{Hierarchical Temporal Memory}

%-----I-----
\acrodef{IAM}{Identity And Access Management}
\acrodef{ICMP}{Internet Control Message Protocol}
\acrodef{IID}{Informally, Identically Distributed}
\acrodef{IoE}{Internet of Everything}
\acrodef{IoT}{Internet of Things}
\acrodef{ITU}{International Telecommunication Union}
\acrodef{IBN}{Intent-Based Networking}
\acrodef{IoA}{Internet of Agents}
\acrodef{ISAC}{Integrated Sensing and Communication}
%-----J-----
%-----K-----
\acrodef{KNN}{K-Nearest Neighbors}
\acrodef{KPI}{Key Performance Indicator}
\acrodef{KPIs}{Key Performance Indicators}
%-----L-----
\acrodef{LSTM}{Long Short-Term Memory}
\acrodef{LLM}{Large Language Model}
%-----M-----
\acrodef{MAE}{Mean Absolute Error}
\acrodef{ML}{Machine Learning}
\acrodef{MLaaS}{Machine Learning as a Service}
\acrodef{MOS}{Mean Opinion Score}
\acrodef{MAPE}{Mean Absolute Percentage Error}
\acrodef{MSE}{Mean Squared Error}
\acrodef{MEC}{Multi-access Edge Computing}
\acrodef{mMTC}{Massive Machine Type Communications}
\acrodef{MFA}{Multi-factor Authentication}
\acrodef{MCP}{Model Context Protocol}
\acrodef{MBR}{Maximum Bit Rate}
\acrodef{MAS}{Multi-Agent System}
%-----N-----
\acrodef{NWDAF}{Network Data Analytics Function}
\acrodef{Net4AI}{Networking for \ac{AI}}
\acrodef{NS}{Network Slicing}
\acrodef{NIC}{Network Interface Card}
%-----O-----
\acrodef{OSM}{Open Source MANO}
%-----P-----
\acrodef{PCA}{Principal Component Analysis}
\acrodef{PFCP}{Packet Forwarding Control Protocol}
\acrodef{PoC}{Proof of Concept}
\acrodef{PPV}{Positive Predictive Value}
\acrodef{PRB}{Physical Resource Block}
\acrodef{PDR}{Packet Detection Rule}
\acrodef{PPO}{Proximal Policy Optimization}
%-----Q-----
\acrodef{QoE}{Quality of experience}
\acrodef{QoS}{Quality of Service}
\acrodef{QER}{Quality of Service Enforcement Rule}
%-----R-----
\acrodef{RAM}{Random Access Memory}
\acrodef{RF}{Random Forest}
\acrodef{RL}{Reinforcement Learning}
\acrodef{RMSE}{Root Mean Square Error}
\acrodef{RNN}{Recurrent Neural Network}
\acrodef{RTT}{Round-Trip Time}
\acrodef{RAN}{Radio Access Network}
\acrodef{RIC}{RAN Intelligent Controller}

%-----S-----
\acrodef{SDN}{Software-Defined Networking}
\acrodef{SFI2}{Slicing Future Internet Infrastructures}
\acrodef{SLA}{Service-Level Agreement}
\acrodef{SON}{Self-Organizing Network}
\acrodef{SMF}{Session Management Function}
\acrodef{S-NSSAI}{Single Network Slice Selection Assistance Information}
\acrodef{SATSI}{Space-Air-Terrestrial-Sea Integrated Networks}

%-----T-----
\acrodef{TQFL}{Trust Deep Q-learning Federated Learning}
\acrodef{TEID}{Tunnel Endpoint Identifier}
\acrodef{TPS}{Transactions Per Second}
\acrodef{TPR}{True Positive Rate}
%-----U-----
\acrodef{UE}{User Equipment}
\acrodef{UEs}{User Equipments}
\acrodef{UPF}{User Plane Function}
\acrodef{UPFs}{User Plane Functions}
\acrodef{PDU}{Packet Data Unit}
\acrodef{URLLC}{Ultra-Reliable and Low Latency Communications}
\acrodef{UAV}{Unmanned Aerial Vehicle}
\acrodef{UAVs}{Unmanned Aerial Vehicles}
\acrodef{UDP}{User Datagram Protocol}
%-----V-----
\acrodef{VoD}{Video on Demand}
\acrodef{VR}{Virtual Reality}
\acrodef{VNF}{Virtual Network Function}
\acrodef{V2X}{Vehicle-to-Everything}

%-----W-----
%-----X-----
\acrodef{XR}{Extended Reality}
%-----Y-----
%-----Z-----
\acrodef{ZTN}{Zero-Touch Network}

\begingroup
\renewcommand\thefootnote{}
\footnotetext{\scriptsize
© 2026 IEEE. Personal use of this material is permitted. Permission from IEEE must be obtained for all other uses, in any current or future media, including reprinting/republishing this material for advertising or promotional purposes, creating new collective works, for resale or redistribution to servers or lists, or reuse of any copyrighted component of this work in other works.

Accepted author manuscript for publication in the Proceedings of the 2027 IEEE Consumer Communications \& Networking Conference (CCNC 2027), Las Vegas, Nevada, USA, January 9--12, 2027.
}
\addtocounter{footnote}{-1}
\endgroup

\begin{abstract}
5G edge network slicing must reconcile two coupled requirements: understanding high-level operator intent and reallocating scarce computing and bandwidth resources under dynamic contention. Centralized optimization can face computational and responsiveness challenges under rapidly changing edge contention, and Multi-Agent System (MAS)-based auctions can support resource arbitration but still depend on operator-defined economic parameters and priorities. Embedding a Large Language Model (LLM) directly in a recurring control loop, by contrast, may introduce non-negligible inference latency and non-deterministic behavior. This paper proposes a two-timescale architecture that addresses this gap: an LLM-based Strategist translates natural-language network intents into economic policy parameters (budget, income, and bidding-aggressiveness) at the macro scale, while a deterministic, Vickrey-inspired second-price mechanism run by SPADE-based slice agents resolves recurring resource contention at the operational scale, with the resulting allocations enforced by patching Kubernetes CPU limits and Calico CNI bandwidth annotations directly on User Plane Function pods. We evaluate the architecture end-to-end on a real Kubernetes/Open5GS edge cluster across two complementary scenarios, each repeated 30 times: a Smart-City scenario, in which natural-language public-safety intents propagate into measurable changes in slice bids and bandwidth allocations within a few auction rounds, and a Hierarchical-slicing scenario, in which the mechanism provides priority-dependent resource and QoE differentiation, adapts to abrupt demand changes, and preserves configured minimum guarantees while maintaining near-complete allocation of the shared bandwidth pool.
\end{abstract}


\section{Introduction}\label{sec:introduction}

The evolution of \ac{5G} networks has intensified the need for mission-critical applications operating at the network edge (\textit{Edge Computing}), where strict latency bounds and reliable access to bandwidth resources are essential to preserve \ac{QoS}. In scenarios such as remote medical procedures or industrial control systems, \textit{Network Slicing} creates isolated logical networks over shared infrastructure and enables differentiated resource allocation, ensuring that sensitive applications retain reserved or prioritized access to resources even under high demand or noisy neighbors~\cite{10173493, 11366571}.
 
Orchestrating these slices at the edge in a distributed scenario poses more challenges than in the network core~\cite{MOREIRA2026107971}. Unlike the core, the edge has scarce and highly contended computational resources (i.e., CPU, memory, and bandwidth). Centralized management approaches can face computational and responsiveness challenges when multiple resource-contention events occur simultaneously~\cite{Andrade2025}. A second, more subtle problem compounds this: even a system that reacts quickly to contention still needs to know \emph{what to prioritize}. Today, the mapping from a high-level operator intent -- a live event that demands more bandwidth, a maintenance window, or an emergency response -- to the concrete resource weight that a slice should receive is largely manual, performed by operators, or hard-coded into static rules that cannot adapt to intents unforeseen at design time.
 
In this context, \textit{Multi-Agent Systems} (MAS) emerge as a promising approach to the first problem. The MAS paradigm distributes the network's decision-making: each slice is represented by an autonomous agent capable of participating in resource negotiation and contention resolution through coordinated interactions with other agents. This approach aligns with the \textit{Self-Organizing Networks} concept, enabling distributed and adaptive responses to abrupt load variations. Yet auction-based MAS coordination still depends on operator-defined economic parameters and priorities. Without an adaptive mechanism for updating these parameters, they can remain as static as the heuristics they replace, effectively shifting the intent-to-infrastructure gap one layer up the control stack.
 
This paper addresses this gap through a two-timescale architecture that separates semantic intent interpretation from deterministic resource arbitration. At the macro scale, a Large Language Model (LLM) acts as a \textit{Strategist}, interpreting natural-language network intents and translating them into economic policy parameters, including slice budget, income rate, and bidding-aggressiveness weight. At the operational scale, these policies parameterize a deterministic Vickrey-inspired second-price mechanism executed by the MAS, in which slice agents compete for available resources according to their economic parameters and observed resource conditions. The resulting allocations are enforced directly on the infrastructure by updating CPU limits and Calico CNI bandwidth annotations on the User Plane Function (UPF) pods of a real Kubernetes/Open5GS edge cluster, without requiring manual intervention in each allocation round. The key design principle is that generative reasoning determines how operator intent influences service priorities, while recurring resource-allocation decisions remain independent of LLM inference.
 
The main contributions of this paper are:
\begin{itemize}
    \item A two-timescale orchestration architecture that separates LLM-based semantic intent translation from deterministic MAS-based resource arbitration, keeping generative reasoning outside the recurring operational control loop;
    \item An intent-to-economic-policy mechanism that translates high-level operator intents into slice-specific priority, budget, income, and bidding-aggressiveness parameters consumed by autonomous slice agents;
    \item An end-to-end implementation using SPADE, Prometheus, Kubernetes, Open5GS, and Calico, with resource-allocation decisions enforced directly on live UPF pods through dynamic CPU and bandwidth updates;    
    \item An experimental evaluation on a real 5G edge testbed across two complementary scenarios, each repeated 30 times, demonstrating intent-driven resource differentiation, adaptation to changing demand, preservation of configured minimum guarantees, and corresponding effects on application-level QoE.
\end{itemize}
This paper is organized as follows. Section~\ref{sec:related_work} surveys related approaches to network-slice orchestration and AI-driven control. Section~\ref{sec:proposed_method} presents the proposed method. Section~\ref{sec:results} reports results and discussion. Section~\ref{sec:concluding_remarks} concludes.

\section{Background \& Related Work}\label{sec:related_work}

\subsection{MAS}

A Multi-Agent System (MAS) is a computational paradigm composed of autonomous, communicating software entities, referred to as agents, that perceive their environment, reason over local objectives, and coordinate with peers to solve problems that benefit from distributed decision-making. Representing each network slice as an autonomous agent allows slice-level state and resource requirements to be expressed independently, while coordination mechanisms arbitrate contention over shared resources.

\subsection{Edge Resource Orchestration and AI-Driven Control}

\textbf{Traditional optimization and heuristics.} Resource orchestration for 5G edge network slicing has been extensively studied through centralized optimization frameworks and heuristic approaches. Mixed-Integer Linear Programming (MILP) has been extensively used to optimize Virtual Network Function (VNF) embedding and multi-domain resource allocation, aiming to minimize operational costs while satisfying capacity constraints \cite{ejaz2025comprehensive, jalalian2023network}. Concurrently, standardized Software-Defined Networking (SDN) and Network Function Virtualization (NFV) orchestrators, such as Open Source MANO (OSM) and ONAP, provide end-to-end slice lifecycle management across heterogeneous domains. At the cloud-native execution layer, Kubernetes-based platforms can be extended with Container Network Interface (CNI) mechanisms to provide tenant isolation and traffic shaping for network slices~\cite{buning2026architecting}.

Despite their strengths, centralized orchestration approaches can face computational and responsiveness challenges as the optimization problem and number of managed resources grow. MILP-based formulations, for example, may incur substantial computational complexity for dynamic resource-allocation problems~\cite{ejaz2025comprehensive}. Conventional orchestration and scheduling mechanisms typically operate on structured telemetry and predefined policies, providing limited support for directly interpreting high-level natural-language operator intents. This creates a complementary challenge for edge orchestration: resource decisions must remain responsive to changing infrastructure conditions while also reflecting service priorities expressed at a higher semantic level.

\textbf{AI-driven and intent-based orchestration.} AI-driven approaches have been widely explored to adapt network-slicing policies and resource allocation to changing conditions. Deep Reinforcement Learning (DRL) and Multi-Agent RL (MARL) can dynamically optimize resource distribution, but training requirements, reward-design complexity, and variability in learned policies can complicate predictable SLA enforcement~\cite{cui2026marl}. More recently, researchers have investigated Large Language Models (LLMs) in Intent-Based Networking (IBN) as semantic interfaces for translating natural-language operator intents into structured network policies~\cite{mekrache2024intent, phone2026performance}. LLM-based intent translation may also introduce additional inference overhead compared with lighter models~\cite{phone2026performance}, which becomes relevant for frequently executed control decisions. These characteristics motivate our two-timescale architecture: an event-driven LLM Strategist translates operator intent into economic policy parameters, while recurring resource arbitration is delegated to a deterministic MAS-based mechanism using current infrastructure telemetry.

\section{Proposed Method}\label{sec:proposed_method}

\begin{figure*}[htbp]
    \centering
    % trim = esquerda inferior direita superior
    % Exemplo: cortando 0cm (esq), 0cm (inf), 2cm (dir) e 1.5cm (topo)
    \includegraphics[trim=0cm 0cm 2.0cm 1.80cm, clip, width=0.9\textwidth]{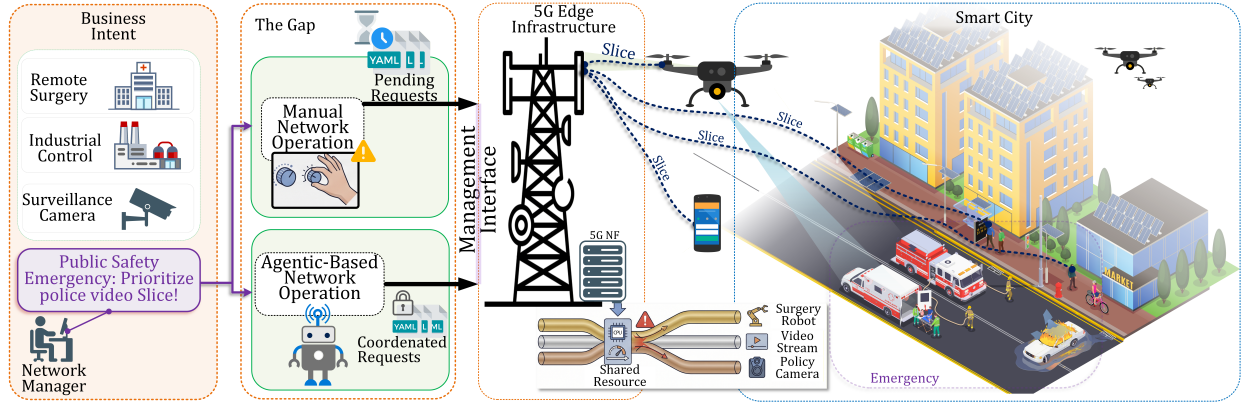}
    \caption{Problem Positioning.}
    \label{fig:problem_positioning}
\end{figure*}

As illustrated in Fig.~\ref{fig:problem_positioning}, the challenge is not merely how to allocate resources for \ac{5G} network slices, but how to translate high-level operator intents into concrete infrastructure actions. High-level intents are commonly realized through manual or static configurations, while resource orchestration must arbitrate dynamic contention over limited computing and bandwidth resources. Our architecture addresses this gap through a two-timescale separation between semantic interpretation and operational resource arbitration. At the macro scale, a \ac{LLM}-based Strategist interprets natural-language intents and translates them into economic policy parameters that express the relative priority of each slice. At the operational scale, autonomous \ac{MAS} agents combine these parameters with current infrastructure telemetry to resolve resource contention through deterministic arbitration. In this way, operator intent influences service priorities without placing generative reasoning inside the recurring resource-allocation loop.

\subsection{Implementation and Use Cases}
% Outra figura que mostre a implementação e já introduza os experimentos
\subsubsection{System Architecture}

The proposed system comprises four cooperating agent roles, implemented with the SPADE framework -- specifically its SPADE-LLM extension for LLM integration -- and communicating asynchronously over XMPP using FIPA-inspired performatives (\texttt{cfp}, \texttt{propose}, \texttt{accept-proposal}, \texttt{reject-proposal}, \texttt{inform}). A \emph{Strategist} agent uses the gpt-oss-20b \ac{LLM} and interprets natural-language network intents. A lightweight \emph{Dispatcher} agent parses the Strategist's structured output and routes individual policy updates to the corresponding \emph{Slice} agents, one per network slice, each bound to a dedicated User Plane Function (UPF) pod. Periodically, a \emph{Resource} agent acts as auctioneer: it runs a sealed-bid auction to resolve contention over the shared CPU and bandwidth pool, and enforces the outcome directly on the underlying Kubernetes cluster. This separation confines the \ac{LLM}'s comparatively slow and non-deterministic reasoning to event-driven policy updates triggered by changes in operator intent, while recurring resource-arbitration decisions are executed deterministically by the \ac{MAS} without further language-model inference.

\subsubsection{Intent-to-Policy Translation}\label{sec:intent_policy}

The Strategist agent is configured with a system prompt that describes the network's slices, their default priority weights, and an explicit economic engine: for a target weight $w \in [1,10]$, the wallet capacity, wealth-regeneration rate, and base bidding aggressiveness of a slice are respectively $B_0 \cdot w$, $I_0 \cdot w$, and $\beta_0 \cdot w$, where $B_0$, $I_0$, and $\beta_0$ are operator-defined base constants. Given a natural-language intent (e.g., an operator reporting a public-safety emergency, as in Fig.~\ref{fig:problem_positioning}), the \ac{LLM} is constrained, via structured output, to return one JSON object per affected slice containing the target slice, its new priority weight, and a short justification. The Dispatcher agent validates and forwards each update as an \texttt{inform} message to the corresponding Slice agent, which immediately recomputes its budget, income, and base bidding aggressiveness from the new weight. Once these parameters are updated, the operator intent is encoded into the economic leverage carried by the slice in subsequent auction rounds, which proceed independently of further LLM inference until a new intent is received.

\subsubsection{Dynamic Bidding and Vickrey-Inspired Resource Arbitration}

Each Slice agent continuously tracks its UPF's CPU, memory, and bandwidth utilization via Prometheus. Upon receiving a call-for-proposals (\texttt{cfp}) from the Resource agent, it maps its highest resource utilization into one of four stress tiers (\emph{high}, \emph{moderate}, \emph{comfortable}, \emph{low}) using configurable thresholds, and computes a priority-weighted coefficient that determines how aggressively it requests additional capacity. Priority $p \in [1,10]$ is normalized and passed through a cubic curve before being mapped into the tier's coefficient range $[c_{min}, c_{max}]$ (as Eq.~\ref{eq:coeff}):

\begin{equation}
\begin{split}
\phi(p) &= \frac{1}{2}\left(\left(\frac{p-5}{5}\right)^{3} + 1\right), \\
c(p) &= c_{min} + \phi(p)\,(c_{max}-c_{min})
\end{split}
\label{eq:coeff}
\end{equation}

so that, under the same congestion level, high-priority slices request larger capacity increments, while low-priority slices remain closer to the lower bound of the coefficient range. The requested target for each resource is $\text{limit} + \text{base}\cdot c(p)$, and the bid itself is $b = \min(\text{budget},\ \beta \cdot m)$, where $\beta$ is the slice's current base bidding-aggressiveness parameter (already scaled by the Strategist-assigned weight from Section~\ref{sec:intent_policy}) and $m$ is a stress multiplier that further amplifies bidding under high congestion and dampens it when the slice is comfortable.

The Resource agent collects all proposals within a fixed window and resolves contention using a Vickrey-inspired sealed-bid, second-price mechanism~\cite{vickrey1961counterspeculation}: the highest bidder is selected, while the corresponding price is determined by the second-highest bid. The winner's request is funded first from any free capacity in the shared pool and, if insufficient, proportionally reallocated from the non-winning slices -- never below each slice's configured minimum floor, with any residual shortfall redistributed starting from the least-resourced non-winning slices so that no active slice is driven below its guaranteed operating minimum. Algorithm~\ref{alg:auction} summarizes a single auction round. The outcome is enforced directly on the cluster: CPU requests and limits are updated via an in-place Kubernetes pod resize, and bandwidth is enforced by patching the Calico CNI QoS annotations on the affected UPF pods, closing the loop from bid to infrastructure within the same auction period without operator intervention. Memory utilization contributes to the bidding decision but is not dynamically enforced in the current prototype because updating memory limits in the present deployment would require restarting the target pod and interrupting the network service. Dynamic memory enforcement is therefore left for future work.

\begin{algorithm}[t]
\caption{Single-round auction and resource reallocation}
\label{alg:auction}
\begin{algorithmic}[1]
\Require Slice agents $\mathcal{S}$, shared pool free capacity, per-slice minimum floors
\State Resource agent broadcasts \texttt{cfp} to all $s \in \mathcal{S}$
\For{each $s \in \mathcal{S}$ (in parallel)}
    \State Measure utilization $u_s$ from Prometheus; select stress tier
    \State Compute coefficient $c_s$ via Eq.~\eqref{eq:coeff} for the tier
    \State $\text{target}_s \gets \text{limit}_s + \text{base}_s \cdot c_s$
    \State $b_s \gets \min(\text{budget}_s,\ \beta_s \cdot m_{\text{tier}})$
    \State Reply \texttt{propose} with $(b_s, \text{target}_s, \text{limit}_s)$
\EndFor
\State $i^{*} \gets \arg\max_{s \in \mathcal{S}} b_s$ \Comment{highest-bid slice}
\State $\text{price} \gets \max_{s \neq i^{*}} b_s$ \Comment{second-highest bid}
\State $\text{need} \gets \text{target}_{i^{*}} - \text{limit}_{i^{*}}$
\State Fund $\text{need}$ from the free pool, then proportionally from non-winning slices, respecting each non-winning slice's minimum floor (remaining shortfall covered least-resourced-first)
\State Grant $i^{*}$ the allocated capacity; debit $\text{price}$ from its budget
\State Patch CPU (K8S resize) and bandwidth (Calico annotations) on all affected UPF pods
\end{algorithmic}
\end{algorithm}

\section{Results and Discussion}\label{sec:results}

We evaluate the proposed architecture on a real Kubernetes/Open5GS edge cluster\footnote{Full cluster specification available at \url{https://computernetworks-ufrgs.github.io/k8s-cluster/}.}, whose UPF pods run on an 8-core amd64 node with 8\,GB of RAM and 35\,GB of storage, running Ubuntu 24.04 LTS and Kubernetes v1.34.9, using two complementary experiment families: one exercises the complete intent-to-infrastructure path, while the other isolates the operational resource-arbitration layer. Each experiment lasts 900\,s and was repeated 30 times; the resource auction runs every 30\,s (30 rounds per run), and all figures below are averages across the 30 repetitions unless noted otherwise. The total CPU and bandwidth pools are fixed at 0.8 cores and 35\,Mbps, respectively, shared by three slices, each bound to a DASH video client used as a QoE proxy.

\subsection{Scenario 1: Smart-City Intents}\label{sec:results_cidade}

In this scenario, three public-safety services (police, firefighters, and an ambulance) compete for the shared pool, and four natural-language intents are delivered to the Strategist agent at fixed points in every run, with the corresponding priority updates generated from the supplied intents rather than manually configured.

\begin{table*}[t]
\centering
\footnotesize
\caption{Smart-City scenario -- aggregated statistics over 30 runs (900\,s each)}
\label{tab:results_smartcity}
\begin{tabular}{lccccccc}
\toprule
\textbf{Slice} & \textbf{Bitrate (kbps)} & \textbf{Buffer (s)} & \textbf{Stalls} & \textbf{Quality switches} & \textbf{Mean bid} & \textbf{BW (Mbps)} & \textbf{CPU (cores)} \\
\midrule
Ambulance   & 5828 $\pm$ 3401 & 12.6 & 0.45 & 17.5 & 62.3 & 9.8  & 0.067 \\
Firefighters & 9748 $\pm$ 4381 & 14.1 & 0.57 & 13.0 & 59.7 & 15.0 & 0.105 \\
Police      & 6024 $\pm$ 3462 & 12.7 & 0.56 & 19.5 & 62.0 & 9.9  & 0.067 \\
\bottomrule
\end{tabular}
\end{table*}

Table~\ref{tab:results_smartcity} summarizes the aggregated QoE and resource-allocation statistics. The ranking observed across the full run (firefighters $>$ police $>$ ambulance) does not follow either the peak priority weight a slice ever received or its elevated-priority duration in isolation. Firefighters combine the longest elevated window (420\,s, $t{=}180$--$600$, $\sim$47\% of the run) with the least-contested first two-thirds of it, before police also escalates at $t{=}420$, and end with the highest average bitrate. Police hold priority 9 for a shorter window (280\,s, $t{=}420$--$700$, $\sim$31\%), but benefit from firefighters stepping back down at $t{=}600$, leaving the second half of their own window less contested. The ambulance reaches the single highest priority weight of the experiment (10) but only from $t{=}600$ onward ($\sim$33\% of the run) and, for part of that window, still competes against an elevated police (until $t{=}700$); it ends with the lowest average bitrate of the three despite its peak weight advantage. The allocation outcome therefore depends jointly on a slice's priority history, its measured resource conditions, and the priorities of simultaneously competing slices, rather than on priority alone.
\begin{figure}[t]
    \centering
    \includegraphics[width=\columnwidth]{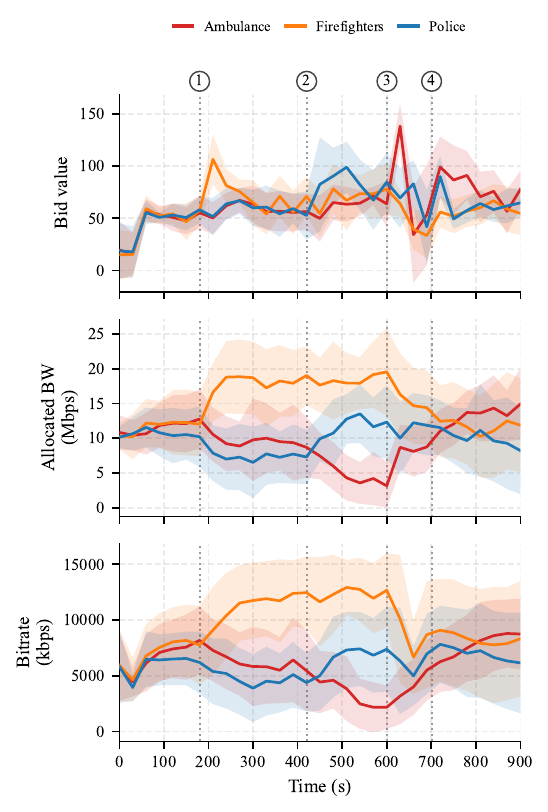}
    \caption{Bid, allocated bandwidth, and bitrate per slice over the four Smart-City intents (mean of 30 runs; shaded band is $\pm 1$ std. dev.). Events: (1) fire report, $t{=}180$\,s, firefighters $\to 9$; (2) police chase, $t{=}420$\,s, police $\to 9$; (3) medical emergency and end of fire, $t{=}600$\,s, ambulance $\to 10$, firefighters $\to 5$; (4) stand-down, $t{=}700$\,s, police $\to 5$.}
    \label{fig:results_smartcity}
\end{figure}

Fig.~\ref{fig:results_smartcity} tracks bid, allocated bandwidth, and bitrate round-by-round across the four intents (30\,s per auction round). Each priority increase produces a bid and bandwidth response within one round, with bitrate lagging by roughly 2--4 rounds (60--120\,s); the corresponding bandwidth reduction falls disproportionately on whichever slice is currently least elevated (event 2 reduces the ambulance's allocated bandwidth by 51\%, while the firefighters' allocation changes only marginally). Event 3 coincides with the DASH content looping back to its lowest representation for all three slices; thus, the immediate bitrate response cannot be attributed solely to resource reallocation, although the bandwidth traces show that the policy changes had already propagated to the infrastructure. Event 4 produces a slower and more variable response than the preceding priority increases. This indicates that the Strategist's intent shapes the allocation outcome without fully determining it: the deterministic resource-arbitration mechanism continues to respond to each slice's measured congestion, consistent with the two-timescale design described in Section~\ref{sec:proposed_method}. The summed allocated bandwidth across the three slices averages 34.7\,Mbps of the 35\,Mbps pool throughout the run, showing that the mechanism allocates nearly the entire configured bandwidth capacity without exceeding the shared-pool limit.

\subsection{Scenario 2: Hierarchical Slices}\label{sec:results_hier}

This scenario isolates the operational resource-arbitration mechanism from the Strategist layer: three slices are assigned static contractual priorities -- Gold (8.0), Silver (5.0), and Bronze (3.0) -- and traffic is paused and resumed on each slice in turn to emulate subscriber disconnection and reconnection, without invoking the Strategist.

\begin{table*}[t]
\centering
\footnotesize
\caption{Hierarchical slicing scenario -- aggregated statistics over 30 runs (900\,s each)}
\label{tab:results_hier}
\begin{tabular}{lccccccc}
\toprule
\textbf{Slice} & \textbf{Bitrate (kbps)} & \textbf{Buffer (s)} & \textbf{Stalls} & \textbf{Quality switches} & \textbf{Mean bid} & \textbf{BW (Mbps)} & \textbf{CPU (cores)} \\
\midrule
Bronze & 3673 $\pm$ 2885  & 11.6 & 0.65 & 19.7 & 25.6 & 5.3  & 0.030 \\
Silver & 4149 $\pm$ 2936  & 12.1 & 0.47 & 12.9 & 32.1 & 9.7  & 0.026 \\
Gold   & 11347 $\pm$ 4477 & 16.0 & 0.09 & 10.6 & 52.8 & 19.7 & 0.088 \\
\bottomrule
\end{tabular}
\end{table*}

Table~\ref{tab:results_hier} shows clear service differentiation across the three priority classes. Gold's average bitrate is 3.09$\times$ Bronze's, exceeding the nominal priority ratio of 8.0/3.0 $\approx$ 2.67$\times$. This amplification is consistent with the cubic priority-to-coefficient mapping $\phi(p)$ in Eq.~\eqref{eq:coeff}, which intentionally strengthens differentiation toward the upper end of the priority range. QoE indicators follow the same ordering: Gold records the lowest stall rate (0.09) and fewest quality switches (10.6), compared with 0.65 stalls and 19.7 switches for Bronze.

\begin{figure}[t]
    \centering
    \includegraphics[width=\columnwidth]{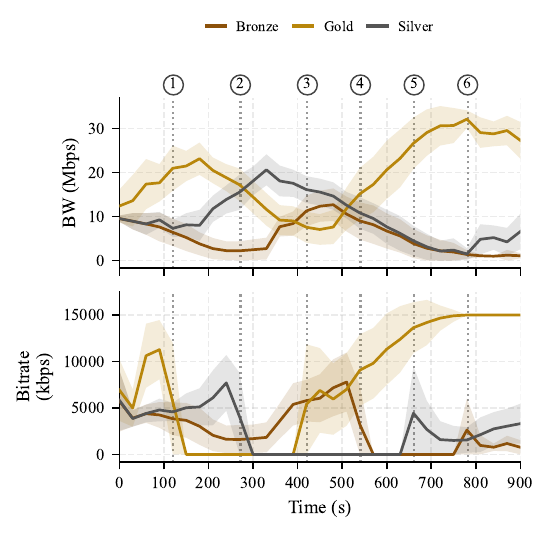}
    \caption{Allocated bandwidth and bitrate per slice (mean of 30 runs; shaded band is $\pm 1$ std. dev.). Events: (1) Gold paused, $t{\approx}121$\,s; (2) Silver paused, $t{\approx}272$\,s; (3) Gold resumed, $t{\approx}422$\,s; (4) Bronze paused, $t{\approx}541$\,s; (5) Silver resumed, $t{\approx}662$\,s; (6) Bronze resumed, $t{\approx}782$\,s.}
    \label{fig:results_hier}
\end{figure}

Fig.~\ref{fig:results_hier} shows the reallocation mechanism directly. Following event 1, Gold's bandwidth falls from about 21.0 to 7.6\,Mbps over 300\,s (ten rounds) as the Resource agent redirects freed capacity to Silver and Bronze; after event 3, it takes about six rounds ($\sim$180\,s) to climb back above 20\,Mbps, tracking the ramp-up of its own measured congestion rather than an instantaneous restoration. Bronze's pause (event 4) drives its allocation toward the configured 1.0\,Mbps floor, and it remains close to that floor even after resuming (event 6), as Gold and Silver's combined demand continues to dominate the pool. Summed bandwidth remains close to the 35\,Mbps pool throughout, including while traffic is simultaneously paused on two slices, showing that configured minimum floors are preserved while the available bandwidth pool is dynamically reassigned as slice demand changes.

\subsection{Discussion}

Taken together, the two scenarios evaluate complementary aspects of the proposed two-timescale architecture. The Smart-City scenario exercises the full intent-to-infrastructure path, showing that natural-language operator intents translate into economic policy parameters and are subsequently reflected in slice bids and bandwidth allocations. The Hierarchical scenario isolates the operational resource-arbitration layer, showing priority-dependent resource and QoE differentiation under changing demand. In both scenarios, the final allocation remains responsive to measured resource conditions rather than being directly prescribed by the LLM, illustrating the intended separation between semantic policy formation and recurring infrastructure control.

%Taken together, the two scenarios validate the paper's two central contributions independently. The Smart-City scenario shows that natural-language business intent can be translated into an actionable economic policy whose effect on infrastructure is observable within a few rounds between an intent being issued and its full effect on delivered bitrate. The Hierarchical scenario shows that, once an economic policy is set, the Vickrey auction resolves contention deterministically and proportionally to priority while remaining resilient to abrupt, unplanned demand changes.

The CPU figures reported in Tables~\ref{tab:results_smartcity} and~\ref{tab:results_hier} (0.03--0.11 cores) reflect the CPU demand observed for the slice workloads during the experiments. The DASH traffic used in both scenarios is primarily bandwidth-driven, which explains the comparatively low CPU demand. These figures do not include the computational cost of the Strategist's LLM inference, which was not instrumented in this deployment. Because the Strategist is invoked only when a new intent arrives rather than in every allocation round, this cost is intermittent and remains outside the recurring resource-arbitration loop. We do not claim the overall architecture has lower computational overhead than a static baseline; quantifying the LLM's inference overhead, and weighing it against the intent-aware differentiation it enables, is left for future work.

\section{Concluding Remarks}\label{sec:concluding_remarks}

This paper presented a two-timescale architecture for autonomous resource orchestration in 5G edge network slicing. An LLM-based Strategist provides a semantic translation layer between natural-language operator intent and slice-specific economic policy parameters, while a deterministic Multi-Agent System performs recurring resource arbitration through a Vickrey-inspired second-price mechanism. This separation combines semantic flexibility at the policy layer with deterministic execution at the operational resource-arbitration layer, without requiring generative inference for each allocation decision. Evaluated end-to-end on a real Kubernetes/Open5GS edge cluster over 30 repetitions per scenario, the Smart-City experiments showed operator intents propagating into measurable changes in slice bids and bandwidth allocations, while the Hierarchical experiments demonstrated priority-dependent resource and QoE differentiation, adaptation to changing demand, and preservation of configured minimum guarantees.

%This paper presented a hybrid, macro/micro architecture for autonomous resource orchestration in 5G edge network slicing, in which a Large Language Model acts as a semantic \emph{Strategist} that translates natural-language business intent into economic policy, while a deterministic Multi-Agent System resolves real-time contention through a sealed-bid Vickrey auction. Confining the LLM's slow, non-deterministic reasoning to an infrequent macro-scale decision, and delegating the millisecond-scale allocation decision to a lightweight, truthful auction, closes the intent-to-infrastructure gap identified in Section~\ref{sec:related_work} without placing generative reasoning inside the time-critical control loop. Validated end-to-end on a real Kubernetes/Open5GS cluster over 30 repetitions per scenario, the Smart-City experiments showed natural-language intents propagating into measurable, bounded infrastructure changes within a few auction rounds, while the Hierarchical experiments showed the auction reallocating bandwidth proportionally to priority, recovering gracefully from abrupt demand changes, and never violating the configured minimum guarantees.

The mechanism's economic parameters were empirically configured for the evaluated scenarios, and their generalization across different slice mixes and traffic patterns remains to be investigated. Future work includes enabling slice agents to learn these parameters through reinforcement learning without interfering with the deterministic resource-arbitration loop~\cite{cui2026marl}, and allowing the Strategist to incorporate utilization telemetry to reconcile operator intents proactively with observed network conditions.

%The auction's economic parameters -- the budget, income, and bid constants, the stress-tier thresholds, and the coefficient range of Eq.~\eqref{eq:coeff} -- were set empirically and are unlikely to generalize across slice mixes or traffic patterns. Future work includes letting each slice agent learn these coefficients online through a reinforcement learning mechanism running at a similar timescale to the Strategist's one, without interfering in the deterministic millisecond-scale auction itself \cite{cui2026marl}, and letting the Strategist also consume the slice agents' utilization telemetry periodically, so that a standing business objective is reconciled with observed network state pre-emptively, rather than only after an operator reports an intent.

\section*{Acknowledgment}

The authors thank FAPESP MCTIC/CGI Research projects 2018/23097-3 - SFI2: Slicing Future Internet Infrastructures and 20/05182-3 - PORVIR-5G: programability, orchestration and virtualization in 5G networks, and Fundação para a Ciência e a Tecnologia (FCT) within the R\&D Unit Project Scope UID/00319/2025 - Centro ALGORITMI (ALGORITMI/UM) for supporting this work.

\bibliographystyle{IEEEtran}
\bibliography{references}

\end{document}